\documentclass
[pla,twocolumn,showpacs,lengthcheck,preprintnumbers]{revtex4-1}%
\usepackage{amsfonts}
\usepackage{amsmath}
\usepackage{amssymb}
\usepackage{graphicx}%
\providecommand{\U}[1]{\protect\rule{.1in}{.1in}}

\begin{document}
\preprint{ }
\title{Persistent currents with non-quantized angular momentum }
\author{A. Mu\~noz Mateo$^1$, A. Gallem\'{i}$^{1,2}$, M. Guilleumas$^{1,2}$, 
  and R. Mayol$^{1,2}$}
\affiliation{$^1$Departament d'Estructura i Constituents de la Mat\`eria,
  Facultat de F\'isica, Universitat de Barcelona, Mart\'i i Franqu\`es, 1, E--08028 Barcelona, Spain 
  \\ $^2$Institut de Nanoci\`encia i Nanotecnologia 
  (IN$\,^2$UB), Universitat de Barcelona }


\begin{abstract}
We analyze the generation of persistent currents in Bose-Einstein 
condensates of ultracold gases confined in a ring. This phenomenon has been 
recently investigated in an experiment [Nature \textbf{506}, 
200 (2014)], where hysteresis 
loops have been observed in the activation of quantized persistent currents by 
rotating weak links. In this work, we demonstrate the existence of 3D 
stationary currents with non-quantized angular momentum. They are generated by 
families of solitary waves that show a continuous variation in the angular 
momentum, and provide a bridge between different winding numbers. 
We show that the size of hysteresis loops is determined by the range of 
existence  within the weak link region of solitary waves which configure the 
energy barrier preventing phase slips. The barrier vanishes when the critical 
rotation leads winding numbers and solitonic states to a matching 
configuration. At this point, Landau and Feynman criteria for phase slips meet:
the fluid flow reaches the local speed of sound, and stationary vortex 
lines (which are the building blocks of solitons) can be excited inside the system.
\end{abstract}
\maketitle

\section{Introduction}
Persistent currents are one of the striking phenomena 
associated to superfluidity. In charged systems, electric currents flowing 
without dissipation along superconductor loops are due to the superfluid flow 
of Bose-condensed Cooper pairs, and have found an extensive field of 
applications since the prediction of SQUIDs \cite{Jaklevic1964}. 
In neutral systems, persistent currents have been observed since the early 
experiments with superfluid liquid helium (see Ref. \cite{Hoskinson2005} and references 
therein), and have experimented a revival with the advent of Bose-Einstein 
condensation of ultracold gases. Multiply connected geometries have 
allowed the generation of long-life currents in Bose-Einstein 
condensates (BECs) 
\cite{Ryu2007,Ramanathan2011,Moulder2012}, paving the way for atomtronic devices.

The effect of a rotating external potential producing a depletion in the 
density (weak link) has been theoretically 
and experimentally addressed in rings of ultracold atomic gases
\cite{Piazza2009,Piazza2013,Ramanathan2011, 
Wright2013}. The rotating weak link drags the superfluid into a moving state
once the resistance of the system to generate excitations has been overcome.
Whether superfluid excitations are triggered by reaching the speed of sound, 
according to Landau criterion, or by vortex nucleation, following Feynman 
criterion, has yet to be elucidated \cite{Piazza2013}. 
Experimental works (see Ref. \cite{Eckel2014} and 
references therein) point out that vortices are responsible of the activation 
of persistent currents in a ring. Phase slips seem to be originated by the
transit of a vortices towards the inner (outer) edge of the system \cite{Abad2015}, 
producing the increase (decrease) up to one unit in the winding number.

In Ref. \cite{Eckel2014} a hysteresis loop has been observed in the generation 
of quantized persistent currents by means of a rotating weak link.
A 3D ring of Bose-condensed $^{23}$Na atoms was prepared in a 
stationary state with winding number $0$ ($1$), and the frequency $\Omega_c^+$ 
($\Omega_c^-$) at which the winding number changes to $1$ ($0$) was measured.
The size of the hysteresis loop, i.e. the difference 
$(\Omega_c^+ -\Omega_c^-)$, 
was compared with numerical results obtained by solving the 3D 
Gross-Pitaevskii equation (GPE), showing a big discrepancy. As the intensity of the 
link decreases, the size of the loop becomes considerably smaller in the 
experiment than in the numerics. The reason adduced for the difference points 
to a dissipation mechanism, but a complete explanation has still not been given.

Hysteresis is a phenomenon intimately linked to systems presenting multiple 
local energy minima separated by energy barriers that must 
be overcome when driving the system from one minimum to the other 
\cite{Mueller2002}. The final 
state depends on the path followed in the dynamical evolution searching for 
equilibrium states. For superfluids in a ring, it is well known that local 
energy minima correspond to states with different winding number 
\cite{Bloch1973}. However, neither the exact nature of the states lying on the 
barrier nor their stability have been shown yet, although it has been 
hypothesized that unstable states separate adjacent winding numbers.

In this work we demonstrate the existence of stationary states with 
non-quantized angular momentum in 3D rings. These states are solitary 
waves that fill the gap of angular momentum between winding numbers. Among 
such solitonic states, there exist dynamically stable configurations 
that can support long-life currents. 
In addition, we show how the rotating weak link provides common features of 
solitary waves to winding number states. In doing so, the size of hysteresis 
loops can be determined by the range of existence of solitary waves within the 
weak link region. Solitonic states configure the energy barrier preventing phase 
slips and meet a family of winding number states when the barrier vanishes. 
This fact indicates that Landau and Feynman criteria may also match. The former stablishes 
that phase slips are produced when the fluid flow reaches the local speed of 
sound, at the end of a winding number family of stationary solutions. But this is also 
in agreement with Feynman criterion, which states that phase slips appear when vortex lines
can be excited inside the system, as it occurs at the end of a 
solitonic family of steady states.

\section{Model} 
We will consider BECs  in a ring-shaped geometry (torus in 3D) of 
radius $R$ rotating at frequency $\Omega$ within the mean-field regime. 
Stationary states 
$\Psi(\mathbf{r},t)=\psi(\mathbf{r}) e^{-i\mu t/\hbar}$  fulfill the 
time-independent GPE (in the comoving frame) with 
chemical potential $\mu$:
 \begin{equation}
  \left(-\frac{\hbar^2}{2m} \mathbf{\nabla}^2   - \Omega \hat{L}_z 
  + V(\mathbf{r})  + g |\psi|^2\right) \psi = \mu \psi \, ,
  \label{3DGP}
  \end{equation}
where $g=4\pi\hbar^2 a/m$ defines the interaction strength as a function of 
the $s$-wave scattering length $a$, and the external potential 
$V(\mathbf{r})$ includes harmonic trapping 
$V_{trap}(\mathbf{r})=m\omega_z^2 z^2 /2 + m \omega_\perp^2 (r_\perp-R)^2 
/2 $, with $r_\perp^2=x^2 +y^2$, and a weak link 
$U_{wl}(\mathbf{r})= U_{0}\exp[-2\, (R(\theta-\theta_0)/w_\theta)^2]$ 
along the azimuthal coordinate $\theta=\arctan(y/x)$. The angular momentum 
operator is defined by $\hat{L}_z=-i\hbar\,\partial_\theta$, and the energy can 
be calculated by $E[\psi] = \int d\mathbf{r} 
\left\{ |(-i\hbar\mathbf{\nabla}-m\mathbf{\Omega}\times\mathbf{r})\psi|^2/2 m +
 V|\psi|^2 + g|\psi|^4/2 \right\}$.

Persistent currents can be supported by states with integer winding number 
$q$, quantized angular momentum per particle 
$\langle\hat{L}_z\rangle/N=\hbar q$, and constant density along $\theta$, that 
is
 \begin{equation}
 \Psi_q(\mathbf{r},t)=e^{-i\mu t/\hbar}\psi_q(z,r_\perp)\,e^{i q\theta} \, .
 \label{vortex}
  \end{equation}
 In addition, persistent currents can also be provided by solitonic states, 
which present non-quantized angular momentum and produce density dips along the 
ring
  \begin{equation}
  \Psi_\kappa(\mathbf{r},t)=e^{-i\mu 
t/\hbar}\psi_\kappa(z,r_\perp,\theta)\,e^{i \kappa 
\theta} \, ,
 \label{soliton}
  \end{equation}
 where $\kappa$, a real number, plays the role of a quasimomentum (see below).
 
 \section{1D rings}
 This framework can be easily visualized in 1D rings. 
 Without weak link, the lowest-energy states for a given angular momentum, 
or \textit{yrast} states, have been identified as solitary waves connecting 
states with different winding numbers 
\cite{Mottelson1999,Kanamoto2009,Jackson2011}. 
Assuming that a tight transverse harmonic trap enforces a 1D geometry, and 
using units of $E_R={\hbar^2}/{mR^2}$, the dimensionless 1D version 
of 
Eq. (\ref{3DGP}) without external potential reads
  \begin{equation}
  \frac{1}{2}\left(-i{\partial_\theta} -\hat{\Omega} \right)^2
  \psi + \hat{g} |\psi|^2 \psi = \hat{\mu} \psi ,
  \label{1DGP}
  \end{equation}
  where 
$\hat{\mu}=\mu/E_R+\hat{\Omega}^2/2$, $\hat{\Omega}
=\hbar\Omega/E_R$ ,  $\hat{g}= 2 a R N/ 
a_\perp^2 $, and $a_\perp=\sqrt{\hbar/m\omega_\perp}$ is the 
characteristic length of the transverse trap.
 \begin{figure}[tb]
  \centering
  \includegraphics[width=0.9\linewidth]{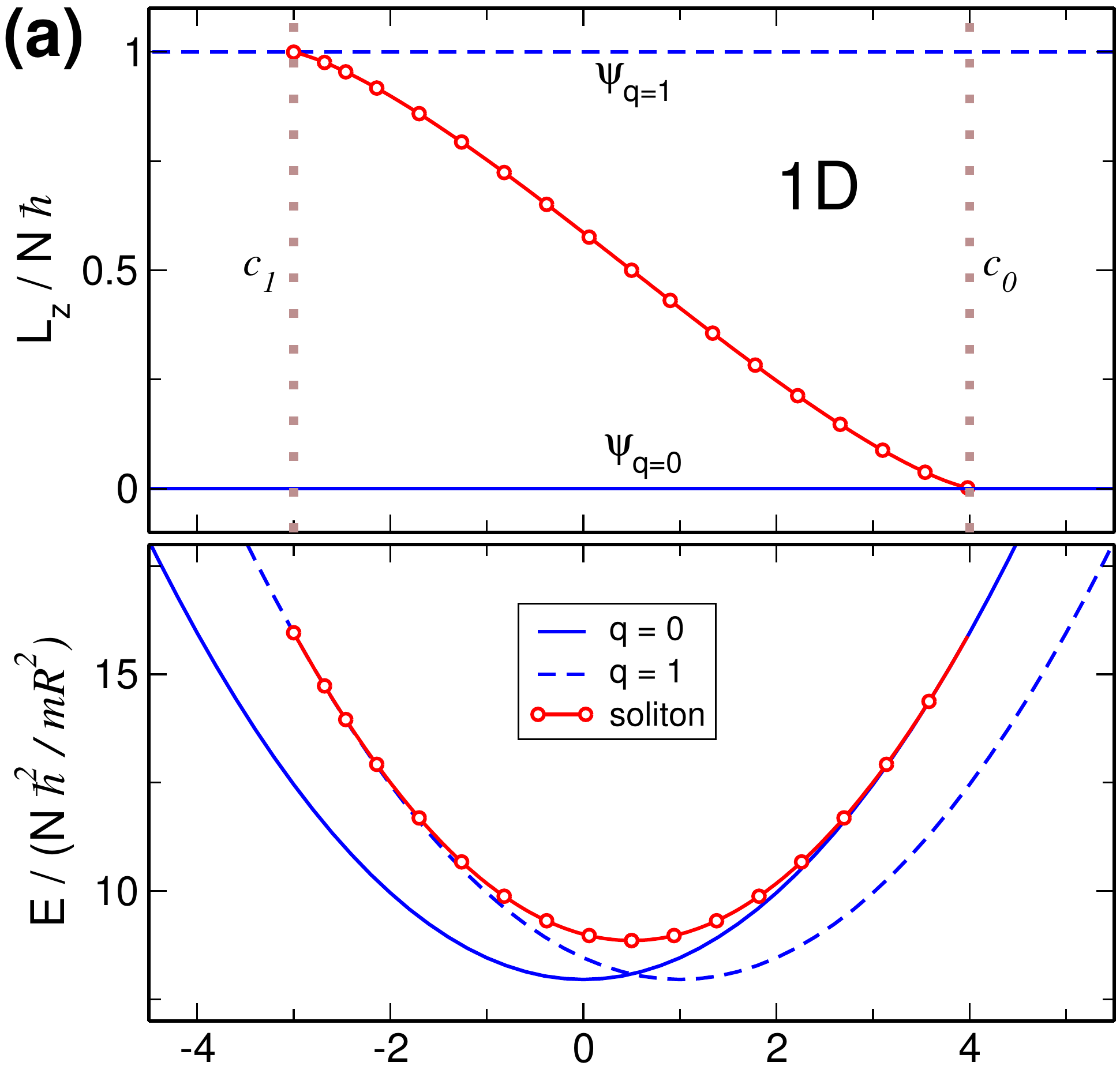}
  \vskip0.2cm
  \includegraphics[width=0.9\linewidth]{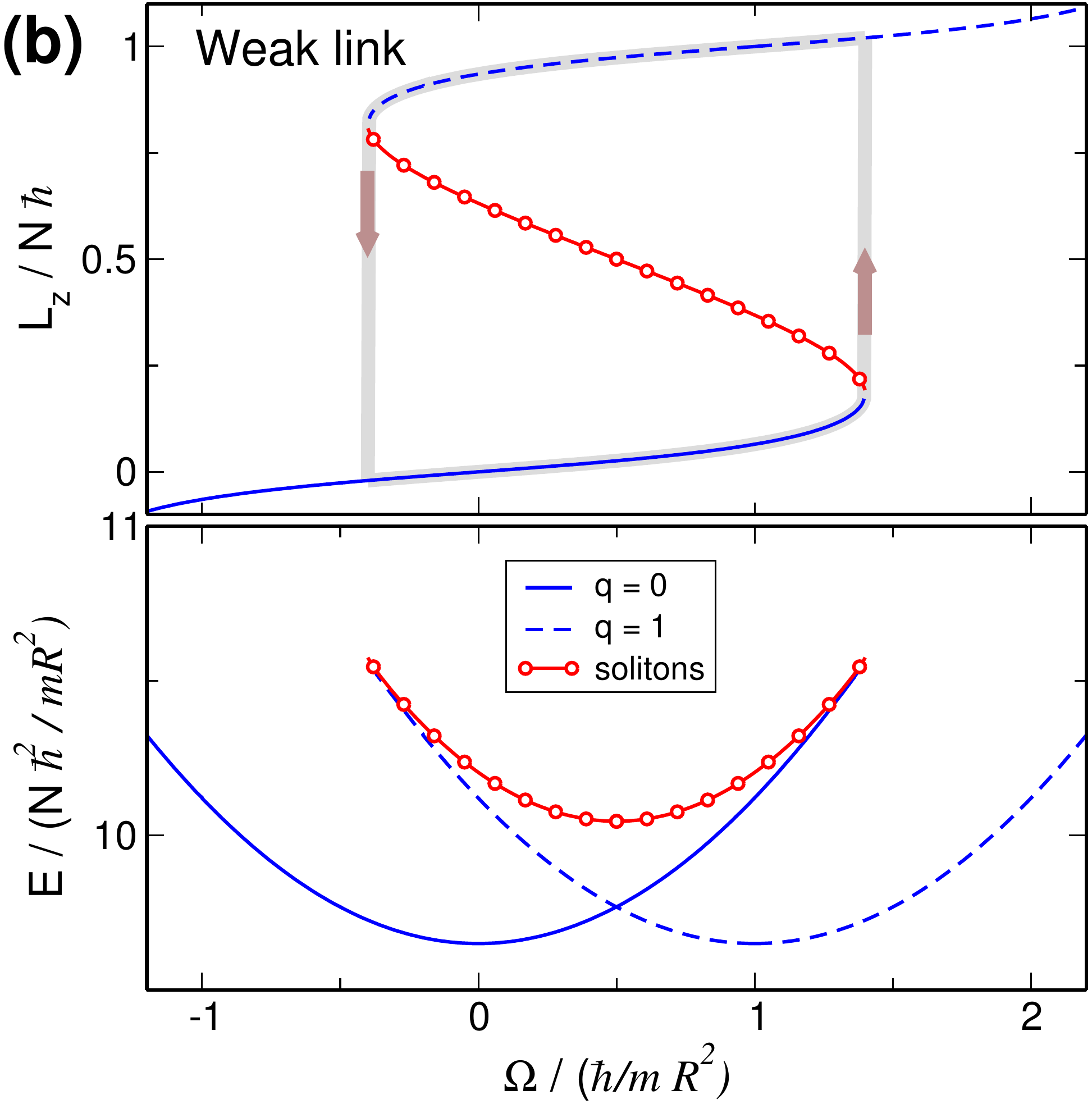}
  \caption{Soliton connection between winding-number states in a 
1D ring for condensates with $\hat{g}=100$. Only winding numbers 
0 and 1 are considered, without (a) and with (b) external weak link of depth 
$U_{wl}=0.56 \,\mu$ and width $w_\theta= \,R$ along the angular 
coordinate. For each 
case, angular momentum per particle is shown above and energy per particle 
below. The thick dotted lines of the top graph (a) indicate the values of the 
speed of sound, $c_0$ and $c_1$, relative to the winding number state 0 (right) 
and 1 (left).
The hysteresis loop is displayed in (b) by thick gray lines and arrows.}
  \label{Fig1Dwinding}
  \end{figure}
Among the analytical solutions of Eq. (\ref{1DGP}), a dark 
soliton moving with linear velocity $v_s$ along the ring is given by Eq. 
(\ref{soliton}) with
      \begin{equation}
   \psi_{\kappa}(\theta) = \sqrt{\frac{\mu}{\hat{g}E_R}}\left[i\beta+\gamma 
\tanh\left(\gamma \sqrt{\frac{\mu}{E_R}}\,\theta\right)\right] \, 
,
  \label{1Dsoliton}
  \end{equation}
 where $\gamma^2+\beta^2=1$, $\beta=v_s/c$, $c=\sqrt{\mu/m}$ is the speed 
of sound, and $\hbar\kappa/m R=\Omega R -v_s $. As can be seen in 
Fig. \ref{Fig1Dwinding}(a), solitons (open circles), given by Eqs. 
(\ref{soliton}) and (\ref{1Dsoliton}), connect states 
$\psi_{q}(\theta)\propto e^{i q \theta}$ having consecutive winding phases 
(solid and dashed lines). 

By producing localized density depletions in the ring, a single soliton 
induces a phase jump $\Delta S= 2 \arccos \beta$ in the condensate wave 
function \cite{Pitaevskii2003}. 
These phase jumps need to be counterbalanced by background 
currents around the torus, with angular speed $\kappa=-q-\Delta S/ 2\pi $, to 
ensure a single-valued order parameter whenever a closed loop is followed. The 
resulting flow carries angular momentum per particle, 
$\langle\hat{L}_z\rangle/N\hbar=-\beta\gamma/\pi-\kappa+(\Delta q-1)/2$, which 
takes continuous values in the range between integer winding numbers $q$ and 
$q+\Delta q$, with $\Delta q=\pm 1$. The velocity of 
moving solitons $v_s$ cannot exceed the speed of sound $c$, at which 
nonlinear soliton excitations transform into linear sound waves traveling on a 
winding number state. Therefore this speed limit bounds the range of existence 
of solitonic solutions. Provided that 
solitonic states can show dynamical stability \cite{Kanamoto2009}, 
their associated flows will appear as persistent currents. We will see 
that there exist the counterpart of these states in multidimensional systems. 

\subsection{Weak link}
The presence of a rotating weak link introduces variations in the density 
profile that give rise to phase jumps, and, like in solitons, 
background currents are generated in order to compensate such jumps. Fig. 
\ref{Fig1Dwinding}(b) displays our results obtained from the numerical solution 
of Eq. (\ref{1DGP}) for a system with $\hat{g}=100$, and weak link of depth 
$U_{wl}=0.56 \,\mu$ and width $w_\theta= \,R$. As can be seen, and 
contrary to the case without weak link, both winding number and solitonic 
states show variations in the angular momentum with the rotation rate.
This effect can be deduced from the expression for the current density 
$\mathbf{j}(\mathbf{r},t)=\rho(\hbar\mathbf{\nabla} S/m- 
\mathbf{\Omega}\times\mathbf{r})$, expressed in terms of density $\rho$ and 
phase $S$ of the order parameter 
$\Psi(\mathbf{r},t)=\sqrt{\rho(\mathbf{r},t)} e^{iS(\mathbf{r},t)}$. Defining 
the velocity as $v={j}/\rho$ we get $ v(\theta)= {\hbar}
(\partial_\theta S(\theta) -\hat{\Omega})/{m R} $ for the 1D 
case. Solving for the phase gradient and integrating along the whole ring we 
obtain
\begin{equation}
2\pi q = 
2\pi\hat{\Omega} +\frac{m R}{\hbar}\oint d\theta \, v(\theta) \, ,
\label{quantization}
\end{equation}
where, due to the single-valuedness of the order parameter, we used 
$S(\theta=2\pi)-S(\theta=0)=2\pi q$. Whenever the 
dimensionless angular rotation $\hat{\Omega}$ takes integer 
values, it is possible to look for static states having $v(\theta)=0$. Instead, 
stirring the system at non-integer $\hat{\Omega}$ values imposes a 
non-vanishing velocity, hence phase gradients.

\begin{figure}[tb]
  \centering
  \includegraphics[width=0.95\linewidth]{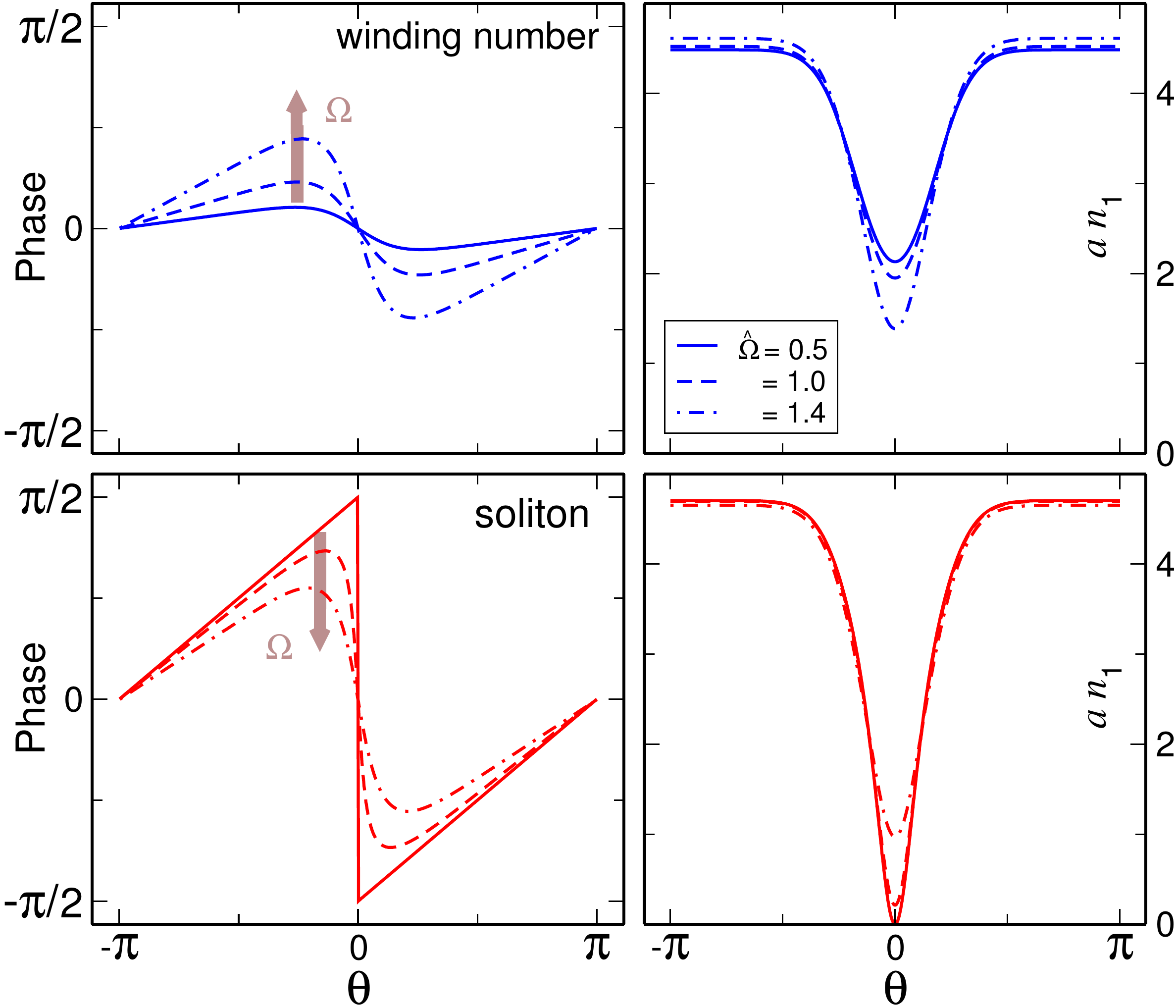}
  \caption{ Phases (left column) and dimensionless azimuthal densities 
  (right column), obtained from numerical solutions of the time-independent 1D 
GPE (\ref{1DGP}), corresponding to winding number $q=0$ (above) and solitonic 
states (below) for the same parameters as in Fig. \ref{Fig1Dwinding}(b),  
rotating at angular frequencies $\hat{\Omega}=$ 0.5, 1.0 and 1.4. 
Thick arrows (on the left column) indicate the 
sequence of both type of states for increasing $\Omega$, approaching convergence
near $\hat{\Omega}=1.4$. }
  \label{Fig1Dphase}
  \end{figure}
  
As a consequence of the weak link, which breaks the rotational 
symmetry of the system, a gap is opened in the energy spectrum of 
winding numbers \cite{Mueller2002}. Our numerical results, displayed in Fig. 
\ref{Fig1Dwinding}(b), quantify the qualitative analysis of this fact made 
previously in the literature. As can be seen, swallow tails are developed in 
the presence of a weak link, cutting the otherwise continuous curves traced 
by winding numbers in Fig. \ref{Fig1Dwinding}(a).  
Generating solitons at the weak link position is energetically more 
favorable than at higher-density regions. For weak link 
extents $w_{\theta}$ greater than the healing length $\xi=\sqrt{\hbar^2/m 
\mu}$, a local speed of sound $c(\theta)=\sqrt{g\, n_1(\theta)/m}$ can be 
defined ($n_1$ being the azimuthal density), which takes the lowest value at the weak link center. As the speed of 
sound is a measure of the range of existence of solitonic solutions, this 
range is reduced inside the low density region, relative to the system without 
weak link.

When rotation is introduced, opposite effects can be observed in solitonic and 
winding-phase states (see Fig. \ref{Fig1Dphase}). For 
increasing angular frequency, solitons decrease their angular momentum, 
whereas winding numbers increase it. In this way, both type of solutions 
converge at some intermediate value, where the superfluid flow reaches 
the speed of sound at the weak link. Therefore, the  bounds for the 
existence of solitons are also applicable to states with a winding phase.

Hysteresis loops, as those observed in the experiment \cite{Eckel2014}, can be 
clearly identified in Fig. \ref{Fig1Dwinding}(b) (thick brown lines and arrows). 
Distinct from previous analyses, our results are able to 
measure the size of the energy barrier preventing phase slips.
Solitonic states are the responsible of such barrier, occupying stationary 
points in the energy landscape of the system. The barrier vanishes when both 
type of states connect, and it is no longer possible to continue beyond these 
points by increasing the angular velocity. As a result the system decays 
to the next lower-energy winding-phase state.
When the intensity of the weak link grows, the speed of sound and hysteresis 
cycle width reduce, and solitons and winding-phase states become energetically 
closer.

\begin{figure}[tb]
  \centering
  \includegraphics[width=0.9\linewidth]{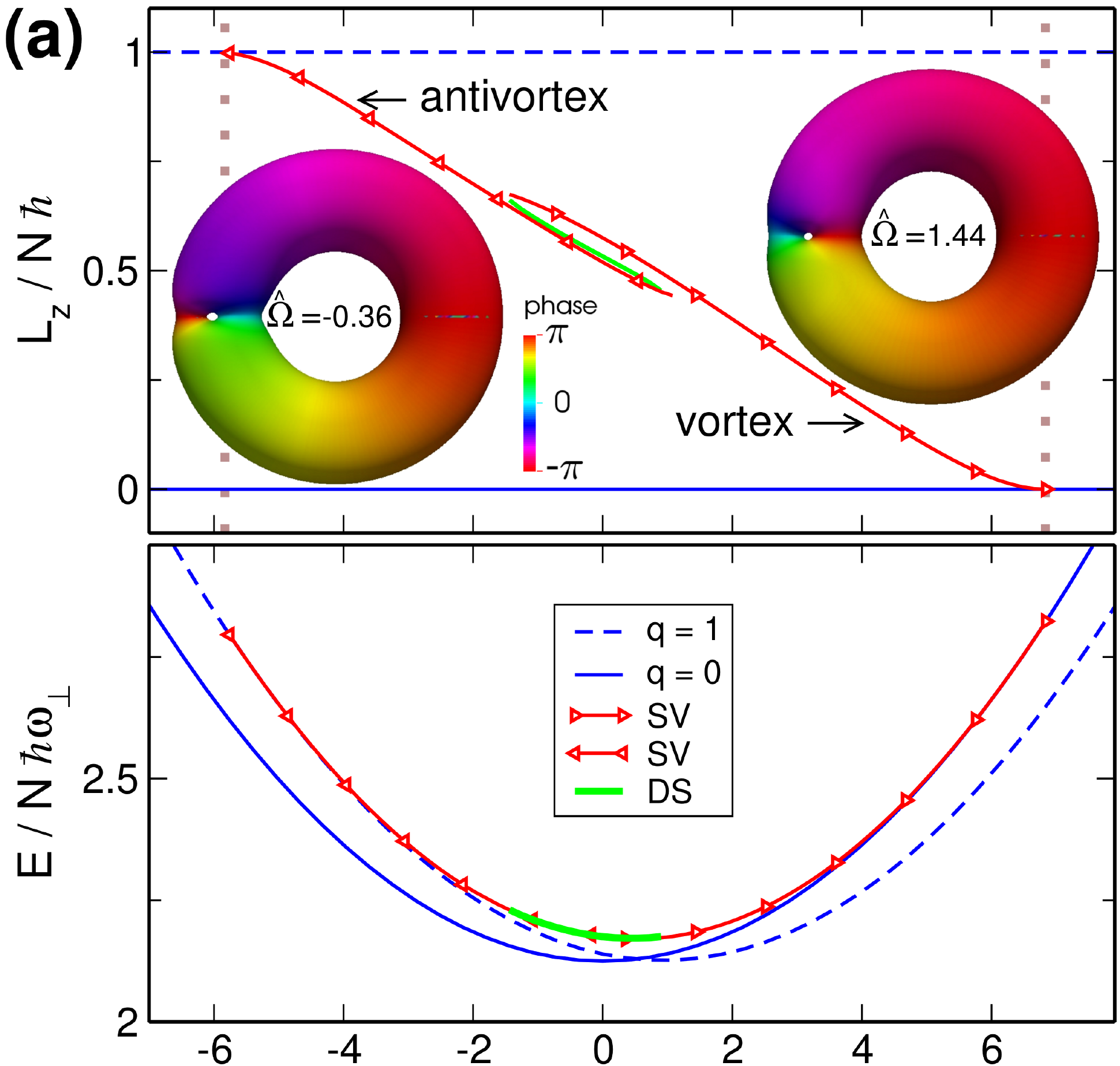}
  \vskip0.2cm
  \includegraphics[width=0.9\linewidth]{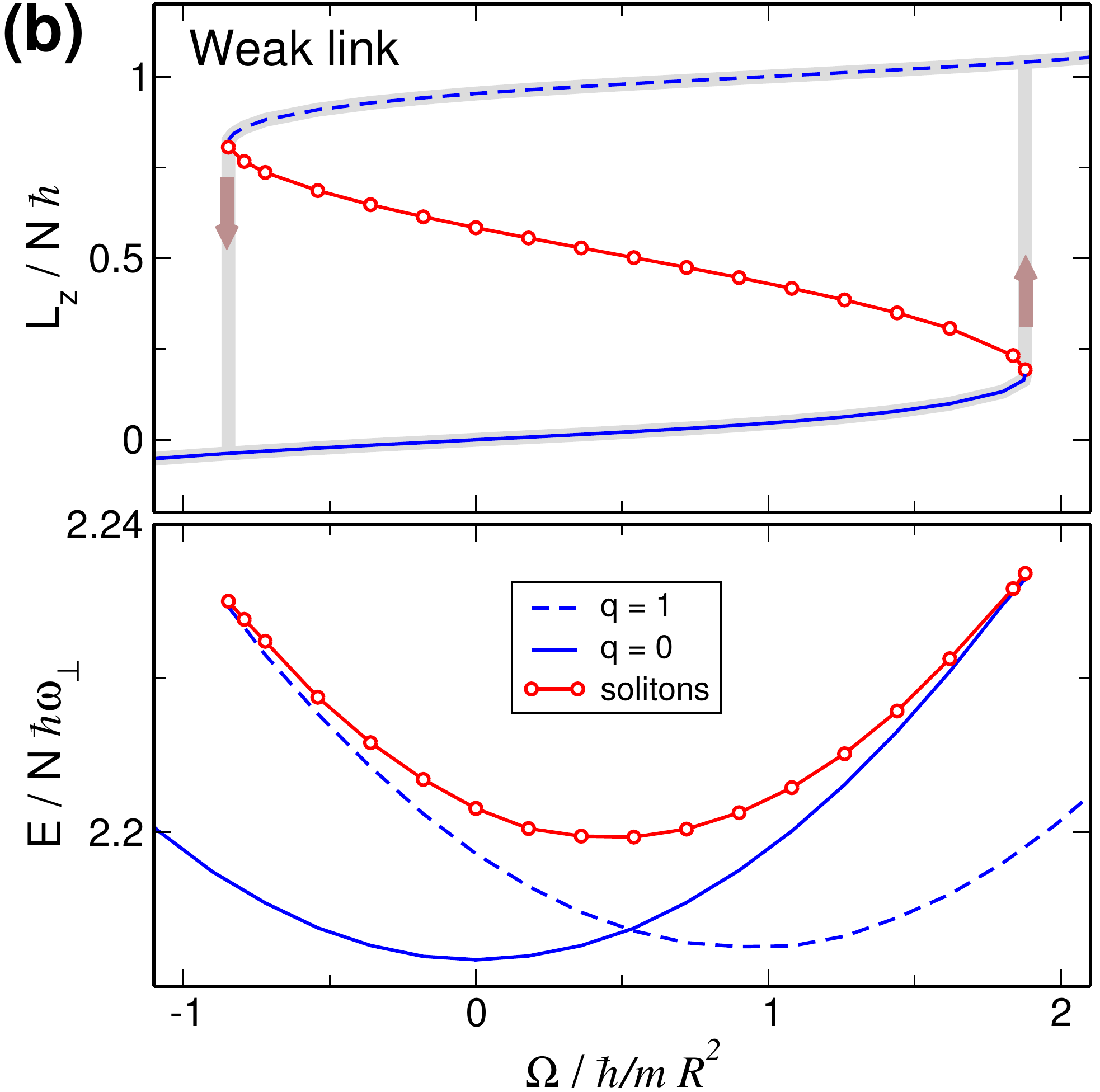}
  \caption{Same as Fig. \ref{Fig1Dwinding} for a 3D system, showing numerical 
results of time-independent 3D GPE (\ref{3DGP}) for condensates containing
$5.3\times \,10^4$ atoms of $^{23}$Na in an isotropic harmonic trap with 
$\omega=2\pi\times100$ Hz centered at $R=6 \,a_\perp$.
Two solitary waves can be excited in the absence of weak link (a), solitonic 
vortex (SV) and dark soliton (DS), whereas 
only a DS is possible at the center of a weak link (b) with depth $U_{wl}=0.5 
\,\mu $ and width $w_\theta= 0.6 \,R$. The insets represent 
density isocontours (at 5$\%$ of maximun density), coloured by phase, of SVs 
belonging to two separated branches containing either vortices or antivortices. 
The external diameter of the tori is $17\,a_\perp$.}
  \label{Fig3Dwinding}
  \end{figure}
\section{3D rings}
In multidimensional systems, the 1D picture outlined before 
is reproduced: solitons are the stationary states making the connection 
between winding numbers and building the energy barrier separating them. 
In addition, dimensionality allows the emergence of novel 
solitonic states and introduces new issues concerning their 
stability. 
It was demonstrated 
in Ref. \cite{MunozMateo2014} that different transverse standing waves can be 
excited in straight-channeled superfluids. A 
dark soliton possesses the simplest structure among such standing waves, and it 
is a dynamically stable configuration for small values of the chemical 
potential, when no other transverse states can be excited. But more complex 
configurations containing vortex lines are also possible. They emerge at higher 
values of the chemical potential, and thus higher densities, and cause the dark 
soliton to turn into a dynamically unstable configuration. The stable state is 
then the solitonic vortex \cite{Brand2001}, i.e. the state containing a single vortex line, 
which have been recently observed in elongated geometries 
\cite{Ku2014,Donadello2014}. It is worth to mention that also in toroidal systems, 
off-centered vortex lines (solitonic vortices) have been observed 
\cite{Wright2013} in the density regions of the condensate.

We have numerically solved the 3D GPE (\ref{3DGP}), by using the Newton method, 
in order to find steady states supporting stationary currents in toroidal 
condensates with typical experimental parameters.
Fig. \ref{Fig3Dwinding}(a) displays data for a regime where, apart from 
winding number states, two types of solitary  waves are possible: dark 
solitons and solitonic vortices. Our results show that 3D toroidal condensates 
support standing waves equivalent to those found in straight channels, and 
follow the same energy criterion for their bifurcation. At the scale of Fig. 
\ref{Fig3Dwinding}(a), the energies of dark solitons and solitonic vortices
are indistinguishable, and the main difference is their range of existence. 
While dark solitons there exist only for values of the angular momentum per 
particle around $(q+1/2)\hbar$, solitonic vortices extend into the regions 
approaching winding numbers. This approach is mediated by the two different 
branches of Fig. \ref{Fig3Dwinding}(a) that are represented by means of 
lines with right and left pointing triangles, along with two 
characteristic density isocontours. The former branch is connecting with the 
winding number $q=0$, and consists of vortices that produce flows rotating in 
the same direction as $\Omega$, while the latter connects with the winding 
number $q=1$ and contains counterrotating vortices (antivortices). 

\begin{figure}[tb]
  \centering
  \includegraphics[width=0.98\linewidth]{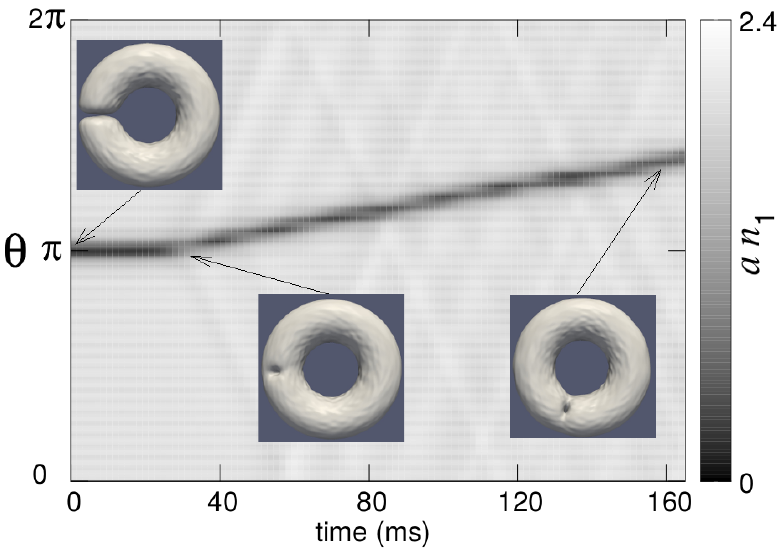}
  \caption{Evolution in real time, obtained by solving the 3D time-dependent 
GPE, of a dark soliton state, with the same parameters as in Fig. 
\ref{Fig3Dwinding}, after an initial Gaussian perturbation. Dimensionless 
azimuthal density $a\hspace{0.1mm}n_1(\theta)$ is plotted as a function of 
time. Density isocontours (at 5$\%$ of maximum density) are showed for times 
(from left to right) 0, 32 and 160 ms. }
  \label{FigDStoSV}
  \end{figure}

It is noteworthy that hysteresis is also present in solitonic vortex states, since 
the two mentioned branches are separated by an energy barrier occupied by dark 
solitons (thick green line in Fig. \ref{Fig3Dwinding}(a)). This energy difference 
between solitonic states is responsible of their stability properties. As shown in 
Fig. \ref{FigDStoSV}, dark solitons decay into solitonic vortices, which are 
dynamically stable states. To show it, we have evolved in real time a dark 
soliton state made of 
$5.3\times \,10^4$ atoms of $^{23}$Na in an isotropic harmonic trap with $\omega=2\pi\times100$ Hz 
centered at $R=6 \,a_\perp$. After adding Gaussian noise to the initial 
stationary solitonic state, we observe that it survives during, approximately, 
the first $20$ ms of the simulation, before decaying into another 
solitary wave, a solitonic vortex, and emitting phonons. 
The subsequent evolution shows the motion of the emergent soliton with 
a different inertial mass than the initial state \cite{Scott2011,Liao2011}.

As in the 1D case, multidimensional solitons can exist only while their 
velocity around the torus is lower than the speed of sound. For isotropic 
harmonic trapping, the speed limit is given by the azimuthal speed of sound 
$c(\theta)$ that can be calculated by \cite{MunozMateo2007}
\begin{equation}
{c}(\theta)=\sqrt{
\frac{{\mu}^2(\theta)-(\hbar\omega_{\perp})^2}{2 m\,{\mu}(\theta)}}
\, .
\label{speedSound}
\end{equation}
In Fig. \ref{Fig3Dwinding}(a), the thick dotted lines represent the values 
given by Eq. (\ref{speedSound}). The manifest agreement with numerical 
solutions of GPE (lines with open triangles), permits Eq. 
(\ref{speedSound}) to be employed for predicting the range where solitons exist 
and hence for measuring the size of hysteresis loops.

\subsection{Weak link}
The presence of the weak link introduces new effects in 3D systems. The 
density depletion reduces not only the local speed of sound, as in 1D 
systems, but also the local energy available for transverse excitations. As a 
result, some solitonic states that can be excited in regions with homogeneous 
density could not survive at the weak link. This is exactly the case 
represented in Fig. \ref{Fig3Dwinding}(b), where solitonic vortices are 
excluded from the weak link region, and only dark solitons can be excited 
inside. As rotation starts, non-null currents are generated in winding-phase 
states. In analogy with the Meissner effect, the system produces a counter flow 
inside the weak link region in order to exclude the vorticity field induced by 
rotation. In a multidimensional system, a vortex-antivortex pair produces such 
a flow \cite{Woo2012,Dubessy2012}. The rotating weak link leads surface modes 
to enter into the torus by nucleating a vortex dipole along the radial 
direction \cite{Piazza2009,Piazza2013}. This dipole yields a phase jump which 
is opposite to the background flow and has the required value for keeping a 
winding phase. The phase jump grows with increasing angular rotation, up to a 
critical point where the flow velocity inside the weak link reaches the local 
speed of sound, and no further increase is allowed. At this point, which is also 
the end of the trajectory for solitons inside the weak link, the localized 
phase jump vanishes and a phase slip is produced. 
This effect is due to the dispersion curve 
of vortex dipoles \cite{Jones1982}, which transform into phonons at low momentum $p$, 
that is to say, when $E/p\rightarrow c$. 
As a consequence, the size of hysteresis loops in 3D rings is 
determined by the range of existence of solitonic solutions, and this range is 
bounded by the local speed of sound $c(\theta)$.

\begin{figure}[tb]
  \centering
  \includegraphics[width=0.95\linewidth]{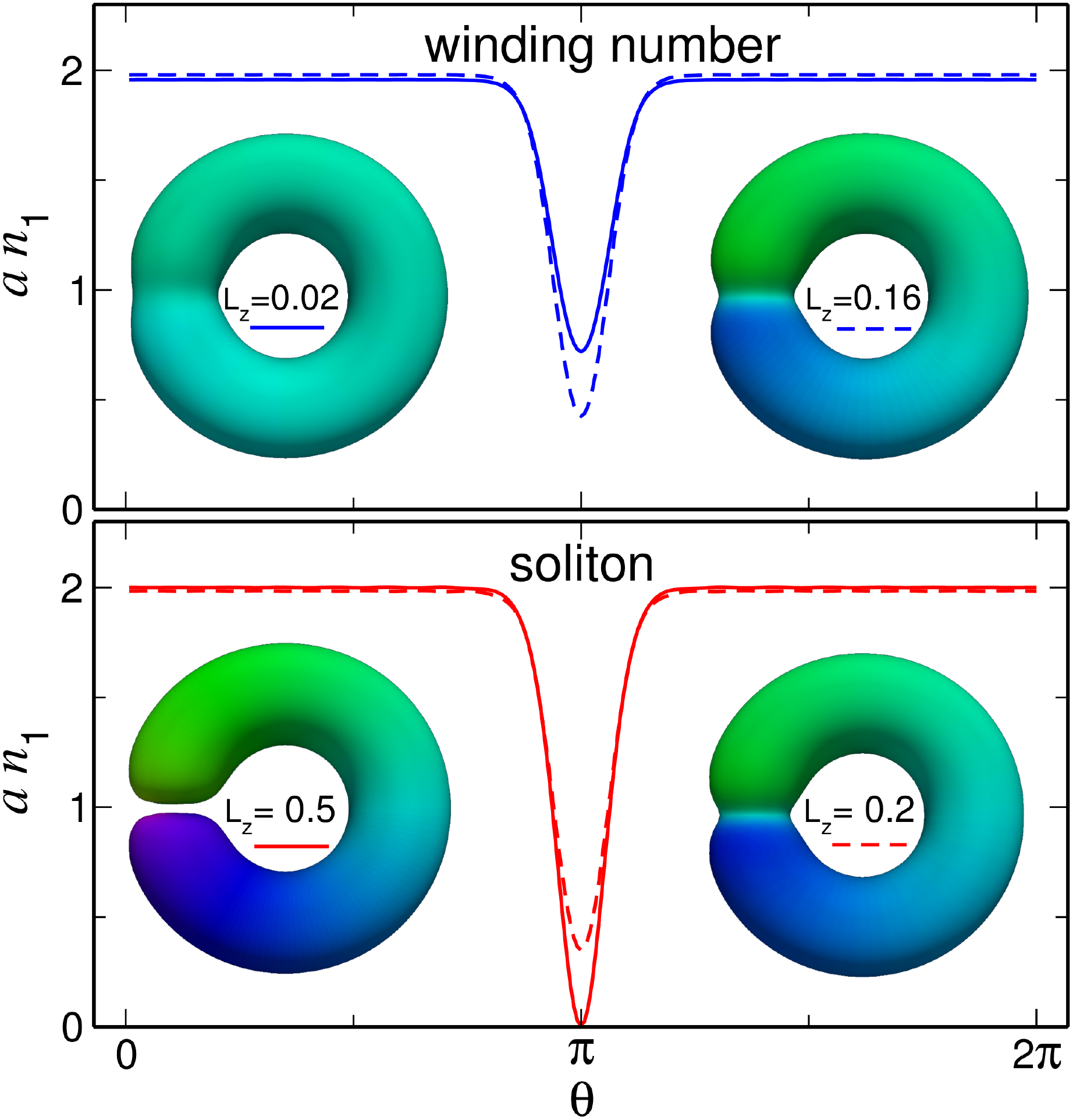}
  \caption{ Density isocontours (5$\%$ of maximum density) of winding number 
$q=0$ (above) and solitonic (below) states, obtained by solving the 
time-independent 3D GPE \ref{3DGP}, in 3D condensates with the same 
trap and rotating weak link as those of Fig. \ref{Fig3Dwinding}. Angular 
momentum per particle $\langle L_z\rangle/N \hbar$ is indicated in the inner 
part of the tori. The angular frequencies considered are $\hat{\Omega}=0.5$ 
(left) and $\hat{\Omega}=1.87$ (right), the latter near the critical frequency 
where states approach $\langle L_z\rangle/N \hbar =0.18$. Background graphs 
show the dimensionless density $an_1(\theta)$, after integrating along 
$(z,r_\perp)$. Colours represent the phase following the colour map of Fig. 
\ref{Fig3Dwinding}. }
  \label{Fig3Dphase}
  \end{figure}
Figure \ref{Fig3Dphase} shows the approach to the critical angular frequency 
followed by winding number states (upper graph in the figure) and solitons 
(lower) for a torus with the same parameters as before. At 
intermediate rotation rates, $\hat{\Omega}=0.5$, a dark soliton has 
angular momentum per particle $\langle L_z\rangle/N \hbar = 0.5$, whereas the 
winding number $q=0$ produces a small current, represented on the figure by 
an almost imperceptible gradient phase. However, as can be seen in the 
isocontours situated on the right at Fig. \ref{Fig3Dphase},  at the rotation 
rate $\hat{\Omega}=1.87$, close to the critical frequency, the differences 
between solitonic and winding numbers begin to disappear, and the 
converging process finishes with a phase slip to winding number $q=1$, when the 
local speed of sound is reached. 

\section{Conclusions}
We have studied the generation of persistent currents by means of rotating 
weak links in 3D Bose-Einstein condensates confined in a torus. We demonstrate 
the existence of persistent currents with non-quantized angular momentum 
supported by stationary states that include solitary waves. They are equivalent 
to the solitonic waves found in straight channels \cite{MunozMateo2014}, and follow 
the same energy criterion for their bifurcation. In the presence of a rotating 
weak link, stationary solitons lying at the weak link give the lowest height of 
the energy barrier separating winding numbers and preventing phase slips. We 
show that the energy barrier vanishes when the angular frequency makes such 
solitons to move at the local speed of sound. At this rotation rate, the 
families of winding numbers and solitons converge, in such a way that 
time-independent states belonging to these families do not exist for a faster 
rotation, and a phase slip is produced. 

Our results point out that the size of the hysteresis loops observed in Ref. 
\cite{Eckel2014} is characterized by the convergence process between winding 
numbers and solitons. This connection 
allows to identify, at the end of the hysteresis loop, both stationary vortex 
lines (making solitonic states), and phonons (traveling at the speed of 
sound on winding number states). This fact indicates that Feynman and Landau 
criteria for the generation of phase slips meet at the connection point. 

However, the 
controversy about the different sizes of hysteresis loops given by 
Gross-Pitaevskii theory and experiments remains.
As long as we are dealing with quantum mechanical systems, quantum tunneling 
processes can be considered in order to get phase slips 
\cite{Freire1997}. It is not necessary 
for the system to wait for changing its angular momentum up to the 
configuration where the energy barrier vanishes. These calculations  
are beyond the aim of the present work and will be reported elsewhere 
\cite{MunozMateo2015}.
%
%
%

\end{document}